\documentclass[twocolumn,showpacs,superscriptaddress]{revtex4}

\usepackage{amssymb}
\usepackage{amsfonts}
\usepackage{amsmath}
\usepackage{graphicx}
\usepackage{mathrsfs}

\begin{document}

\title{Entanglement oscillation and survival induced by non-Markovian decoherence dynamics of entangled squeezed-state}
\author{Jun-Hong An}
\email{phyaj@nus.edu.sg}
\affiliation{Department of Modern Physics, Lanzhou University, Lanzhou 730000, P. R. China}
\affiliation{Centre for Quantum Technologies and Department of Physics, National University of Singapore, 3 Science Drive 2, Singapore 117543, Singapore}
\author{Ye Yeo}
\affiliation{Centre for Quantum Technologies and Department of Physics, National University of Singapore, 3 Science Drive 2, Singapore 117543, Singapore}
\author{Wei-Min Zhang}
\affiliation{Department of Physics and Center for Quantum Information Science, National Cheng Kung University, Tainan 70101, Taiwan}
\author{C. H. Oh}
\email{phyohch@nus.edu.sg}
\affiliation{Centre for Quantum Technologies and Department of Physics, National University of Singapore, 3 Science Drive 2, Singapore 117543, Singapore}
\begin{abstract}
We study the exact decoherence dynamics of the entangled squeezed state of two single-mode optical fields interacting with two independent and
uncorrelated environments.  We analyze in detail the non-Markovian effects on the entanglement evolution of the initially entangled squeezed state for
different environmental correlation time scales.  We find that the environments have dual actions on the system: backaction and dissipation.  In
particular, when the environmental correlation time scale is comparable to the time scale for significant change in the system, the backaction would
counteract the dissipative effect.  Interestingly, this results in the survival of some residual entanglement in the final steady state.
\end{abstract}
 \maketitle

\section{Introduction}
Studies on the decoherence dynamics of open quantum systems are of great importance to the field of quantum information science \cite{Nielsen00}.  Any
realistic analysis of quantum information protocols should take into account the decoherence effect of the environment.  In many quantum communication
and computation schemes, information is transmitted using photons.  For instance, the first experimental verification of quantum teleportation
\cite{Bennett} used pairs of polarization entangled photons to transfer the polarization state of one photon onto another \cite{Bouwmeester97}.
Within a year, unconditional quantum teleportation of optical fields was demonstrated experimentally using squeezed-state entanglement
\cite{Braunstein98, Furusawa98}.  Given their central role in these schemes and many others, much work has been carried out on the decoherence dynamics
of optical fields.  In particular, several authors have studied the continuous variable entanglement of optical fields (see, for instance,
Refs. \cite{Jakub04, An05, Rossi06, Ban06, Goan07, Paris07, An07}).

Conventional approaches not only treat the interactions between the quantum system $S$ of interest and its environment $E$ perturbatively, they also
assume that the environmental correlation time $\tau_E$ is small compared to the time scale $\tau_0$ for significant change in $S$.  These yield
approximate equations of motion, i.e. master equations, under the Born-Markov approximation \cite{Carmichael93, Breuer02}.  Indeed, many studies on the
entanglement dynamics of continuous variable system relied on this approximation \cite{Jakub04, An05, Rossi06}.  However, it is evident from recent
experiments (see, for instance, Refs. \cite{Dubin07, Koppens07, Mogilevtsev08}), that there are many physically relevant situations where the Markovian
assumption does not hold, and a non-Markovian treatment of the open system dynamics is necessary.  So, there has been an increasing interest in the
understanding of the decoherence effect of open quantum system going beyond the Born-Markovian approximation in the last decades \cite{Breuer02, Weiss}.

Very recently, some phenomenological models on non-Markovian entanglement dynamics of optical fields have been investigated \cite{Ban06, Goan07, Paris07}.
It was found that in contrast to the monotonic decrease of entanglement over time in Born-Markovian entanglement dynamics \cite{Jakub04, An05, Rossi06},
there are transient entanglement oscillations in non-Markovian ones.  These oscillations are caused by the backactions of the environments on their
respective local quantum systems \cite{Goan07, Paris07}.  The backaction, characteristic of non-Markovian dynamics, means that the environments with
their states changed due to interactions with the systems, in turn, exert their dynamical influences back on the systems.

In this paper we consider the exact decoherence dynamics of the continuous variable entangled squeezed state of two single-mode optical fields, $S_1$
and $S_2$, that are spatially separated.  Each optical field, $S_k$, interacts with its own environment $E_k$ ($k = 1, 2$).  $E_1$ and $E_2$ are
independent and uncorrelated.  We study the exact entanglement dynamics of the two optical-field system for different $\tau_E$'s in comparsion with
$\tau_0$, and analyze when the system dynamics will exhibit novel non-Markovian effects, and provide a detailed description of these.  To this end, we
use the influence functional formalism \cite{Feynman63, Leggett87}, developed explicitly in Refs. \cite{An07, An072, An08}.  Our results show that besides
the short-time oscillations, the non-Markovian effect can affect the long-time behavior of the system dynamics and the steady state as well.  In
particular, when $\tau_E$ is comparable to $\tau_0$, we find that the backaction effects counteract the dissipative effects of $E_1E_2$ on $S_1S_2$
respectively.  This leads to there being some nonzero residual entanglement in the steady state.

Our paper is organized as follows.  In Sec. II, we introduce a model of two single-mode optical fields in two independent and uncorrelated environments;
and outline the exact dynamics that was derived in detail in Ref. \cite{An08}.  In Sec. III, using logarithmic negativity as an entanglement measure of
continuous-variable states, we discuss the entanglement dynamics of the entangled squeezed state.  Sec. IV presents the numerical results of the
entanglement dynamics, where we analyze explicitly the non-Markovian effect of the environments on the system for different $\tau_E$'s in comparison
with $\tau_0$.  Finally, we conclude in Sec. V.

\section{The total Hamiltonian and exact reduced system decoherence dynamics}
The total Hamiltonian of the system $S_1S_2$ plus environment $E_1E_2$ is given by
\begin{equation}
H = H_{\mathrm{S}} + H_{\mathrm{E}} + H_{\mathrm{I}},
\end{equation}
where
\begin{eqnarray}
H_{\mathrm{S}} & = & \sum_{k = 1}^{2}\hbar\omega_ka_k^{\dag}a_k, \nonumber \\
H_{\mathrm{E}} & = & \sum_{k = 1}^{2}\sum_{l}\hbar\omega_{kl}b_{kl}^{\dag}b_{kl}, \nonumber \\
H_{\mathrm{I}} & = & \sum_{k = 1}^{2}\sum_{l}\hbar(g_{kl}a^{\dag}_kb_{kl} + g_{kl}^{\ast}a_kb_{kl}^{\dag}),
\end{eqnarray}
are, respectively, the Hamiltonian of the two optical fields, the two independent environments, and the interactions between them.  The operators $a_k$
and $a^{\dag}_k$ ($k = 1, 2$) are respectively the annihilation and creation operators of the $k$-th optical mode with frequency $\omega_k$.  The two
independent environments are modeled, as usual, by two sets of harmonic oscillators described by the annihilation and creation operators $b_{kl}$ and
$b^{\dag}_{kl}$.  The coupling constants between the $k$-th optical field and its environment are given by $g_{kl}$.  Currently, most quantum optical
experiments are performed at low temperatures and under vacuum condition.  In this case, vacuum fluctuations are the main source of decoherence.
Therefore, we take the environments to be at zero temperature throughout this paper.

Since we are only interested in the dynamics of $S_1S_2$, we like to eliminate the degrees of freedom of $E_1E_2$.  The influence-functional theory of
Feynman and Vernon \cite{Feynman63} enables us to do that exactly.  By expressing the forward and backward evolution operators of the density matrix of
the system $S_1S_2$ plus environment $E_1E_2$ as a double path integral in the coherent-state representation \cite{Zhang90}, and performing the
integration over the degrees of freedom of $E_1E_2$, we incorporate all the environmental effects on $S_1S_2$ in a functional integral named influence
functional \cite{Feynman63, An07, An08}.  The reduced density matrix, which fully describes the dynamics of $S_1S_2$ is given by
\begin{eqnarray}\label{rout}
\rho (\boldsymbol{\bar{\alpha}}_{f},\boldsymbol{\alpha }_{f}^{\prime };t)
&=&\int d\mu (\boldsymbol{\alpha }_{i})d\mu (\boldsymbol{\alpha }%
_{i}^{\prime })\mathcal{J}(\boldsymbol{\bar{\alpha}}_{f},\boldsymbol{\alpha }%
_{f}^{\prime };t|\boldsymbol{\bar{\alpha}}_{i},\boldsymbol{\alpha }%
_{i}^{\prime };0)  \notag \\
&&~~~~~~~~~\times \rho (\boldsymbol{\bar{\alpha}}_{i},\boldsymbol{\alpha }%
_{i}^{\prime };0),
\end{eqnarray}%
where $\rho(\boldsymbol{\bar{\alpha}}_f,\boldsymbol{\alpha}^{\prime}_f; t) =
\langle\boldsymbol{\alpha}_f|\rho(t)|\boldsymbol{\alpha}^{\prime}_f\rangle$ is the reduced density matrix expressed in coherent-state
representation and
$\mathcal{J}(\boldsymbol{\bar{\alpha}}_f, \boldsymbol{\alpha}^{\prime}_f; t|\boldsymbol{\bar{\alpha}}_i, \boldsymbol{\alpha }^{\prime}_i; 0)$
is the propagating function.  In the derivation of Eq. (\ref{rout}), we have used the coherent-state representation
\begin{equation}
|\boldsymbol{\alpha}\rangle = \prod_{k = 1}^2|\alpha_k\rangle, ~|\alpha_k\rangle = \exp (\alpha _ka^{\dagger}_k)|0_k\rangle.
\end{equation}
which are the eigenstates of annihilation operators, i.e.
$a_k|\alpha_k\rangle = \alpha_k|\alpha_k\rangle$ and obey the
resolution of identity, $\int d\mu \left( \boldsymbol{\alpha
}\right) |\boldsymbol{\alpha }\rangle \langle \boldsymbol{\alpha }|
= 1$ with the integration measures defined as
$d\mu\left(\boldsymbol{\alpha}\right) =
\prod_{l}e^{-\bar{\alpha}_{l}\alpha_{l}}\frac{d\bar{\alpha}
_{l}d\alpha _{l}}{2\pi i}$. $\bar{\boldsymbol{\alpha}}$ denotes the
complex conjugate of $\boldsymbol{\alpha}$.

The time evolution of the reduced density matrix is determined by the propagating function
$\mathcal{J}(\boldsymbol{\bar{\alpha}}_f, \boldsymbol{\alpha}^{\prime}_f; t|\boldsymbol{\bar{\alpha}}_i, \boldsymbol{\alpha}^{\prime}_i; 0)$.  The
propagating function is expressed as the path integral governed by an effective action which consists of the free actions of the forward and backward
propagators of the optical-field system and the influence functional obtained from the integration of environmental degrees of freedom.  After
evaluation of the path integral, the final form of the propagating function is obtained as follows
\begin{eqnarray}
& & \mathcal{J}(\boldsymbol{\bar{\alpha}}_{f},\boldsymbol{\alpha }%
_{f}^{\prime};t|\boldsymbol{\bar{\alpha}}_{i},\boldsymbol{\alpha }%
_{i}^{\prime };0)=\exp \Big\{\sum_{k=1}^{2}\big[u_k(t)\bar{\alpha}%
_{kf}\alpha _{ki}  \notag \\
&&~~~~~+\bar{u}_k(t)\bar{\alpha}_{ki}^{\prime }\alpha _{kf}^{\prime
}+[1-\left\vert u_k(t)\right\vert ^2]\bar{\alpha}_{ki}^{\prime }\alpha _{ki}%
\big]\Big\},  \label{prord}
\end{eqnarray}
where $u_k(\tau)$ satisfies
\begin{equation}\label{ut}
\dot{u}_k(\tau) + i\omega_k u_k(\tau) + \int^{\tau}_0 \mu_k(\tau - \tau^{\prime})u_k(\tau^{\prime}) = 0
\end{equation}
with $\mu_k(x) \equiv \sum_le^{-i\omega_lx}\left\vert g_{kl}\right\vert^2$ being used.  Combining Eq. (\ref{prord}), we can get the exact time-dependent
state from any initial state by the evaluation of the integration in Eq. (\ref{rout}).

To compare with the conventional master equation description of such system, we now derive a master equation from the above results.  After taking the
time derivative to Eq. (\ref{rout}) and recalling the explicit form of Eq. (\ref{prord}), we can derive an exact master
equation
\begin{eqnarray}
\dot{\rho}(t) &=&-\frac{i}{\hbar }[H^{\prime }(t),\rho (t)]+\sum_{k=1} ^2\Gamma_k
(t)[2a_{k}\rho (t)a_{k}^{\dag }  \notag \\
&&-a_{k}^{\dag }a_{k}\rho (t)-\rho (t)a_{k}^{\dag }a_{k}] ,  \label{mas}
\end{eqnarray}
where
\begin{eqnarray}
H^{\prime }(t)=\sum_{k=1} ^2 \hbar \Omega_k (t)a_{k}^{\dag }a_{k},  \label{go}
\end{eqnarray}
is the modified Hamiltonian of the two optical modes and
\begin{equation}
\frac{\dot{u}_k(t)}{u_k(t)}\equiv -\Gamma_k (t)-i\Omega_k (t).  \label{pa}
\end{equation}
Eq. (\ref{mas}) is the exact master equation for the optical-field system.  $\Omega_k(t)$ plays the role of a time-dependent shifted frequency of the
$k$-th optical field.  $\Gamma_k(t)$ represents the corresponding time-dependent decay rate of the field.  We emphasize that the derivation of the
master equation goes beyond the Born-Markovian approximation and contains all the backactions between the system and the environments self-consistently.
All the non-Markovian character resides in the time-dependent coefficients of the exact master equation.

The time-dependent coefficients in the exact master equation, determined by Eq.  (\ref{pa}), essentially depend on the so-called spectral density, which
characterizes the coupling strength of the environment to the system with respect to the frequencies of the environment.  It is defined as
$J_l(\omega)=\sum_{k}\left\vert g_{lk}\right\vert^{2}\delta (\omega -\omega _{l})$.  In the continuum limit the spectral density may have the form
\begin{equation}\label{spectral}
J_k(\omega )=\eta_k \omega \Big( \frac{\omega }{\omega _{c}}\Big)^{n-1} e^{-\frac{\omega }{\omega_{c}}} ,
\end{equation}
where $\omega_{c}$ is an exponential cutoff frequency, and $\eta_k $ is a dimensionless coupling constant between $S_k$ and $E_k$.  The environment is classified as Ohmic if
$n = 1$, sub-Ohmic if $0 < n < 1$, and super-Ohmic for $n > 1$ \cite{Leggett87, Hu92}.  Different spectral densities manifest different non-Markovian
decoherence dynamics.

We note that our exact master equation reduces to the conventional master equation under the relevant Markov approximation.  The coefficients in the
master equation (\ref{mas}) become time-independent \cite{An08}
\begin{eqnarray}\label{mp}
\Gamma_k(t)  & = & \pi J_k (\omega_k),  \notag \\
\Omega_k (t) & = & \omega_k - \mathcal{P}\int^{+\infty}_0\frac{J(\omega)d\omega}{\omega-\omega_k},
\end{eqnarray}
where $\mathcal{P}$ denotes the Cauchy principal value.  The coefficients in Eqs. (\ref{mp}) are precisely the corresponding ones in the Markovian master
equation of the optical system \cite{Carmichael93}.

\section{The non-Markovian entanglement dynamics of the entangled squeezed state}
Initially at time $t = 0$, $S_1S_2$ is in an entangled squeezed state. The entangled two-mode squeezed state is defined as the vacuum state acted on by the two-mode squeezing operator
\begin{equation}\label{initial}
|\psi(0)\rangle = \exp[r(a_1a_2 - a^{\dag}_1a^{\dag}_2)]|00\rangle,
\end{equation}
where $r$ is the squeezing parameter.  In the coherent-state representation, this initial state is given by
\begin{equation}
\rho(\boldsymbol{\bar{\alpha}}_i, \boldsymbol{\alpha}^{\prime}_i; 0) =
\frac{\exp[-\tanh r(\bar{\alpha}_{1i}\bar{\alpha}_{2i} + \alpha^{\prime}_{1i}\alpha^{\prime}_{2i})]}{\cosh^2r}.
\end{equation}
The state approaches the ideal Einstein-Podolsky-Rosen (EPR) state \cite{EPR35} in the limit of infinite squeezing ($r\rightarrow \infty$).
The traditional way to generate the entangled two-mode squeezed state is via the nonlinear optical process of parametric down-conversion \cite{Ou92}.
Recently, a microwave cavity QED-based scheme to generate such states has also been proposed \cite{Pielawa07}.
After generating the entangled state given by Eq. (\ref{initial}), the two cavity fields are then propagated, respectively, to the two locations
separated between the sender and the receiver.  A quantum channel is thus established through the entangled two-mode squeezed state and is ready for
teleporting unknown optical coherent states \cite{Braunstein98,Furusawa98}.

At $t > 0$, due to interactions with $E_1E_2$, $|\psi(0)\rangle$ evolves to a mixed state. A straightforward way to obtain the time-dependent
mixed state is by integrating the propagating function over the initial state of Eq. (\ref{rout}). Then the time-evolution solution of the reduced
density matrix can be obtained exactly,
\begin{equation}
\rho (\boldsymbol{\bar{\alpha}}_{f},\boldsymbol{\alpha }_{f}^{\prime
};t)=a\exp [\sum_{k\neq k^{\prime }}(\frac{b}{2}\bar{\alpha}_{kf}\bar{\alpha}%
_{k^{\prime }f}+c\bar{\alpha}_{kf}\alpha _{kf}^{\prime }+\frac{b^{\ast }}{2}%
\alpha _{kf}^{\prime }\alpha _{k^{\prime }f}^{\prime })],  \label{final}
\end{equation}
where
\begin{eqnarray}
a & = & \frac{1}{\cosh ^{2}\left\vert r\right\vert [1-\tanh ^{2}\left\vert r\right\vert (1-\left\vert u(t)\right\vert ^{2})^{2}]}, \\
b & = & \frac{-\tanh \left\vert r\right\vert u(t)^{2}}{1-\tanh ^{2}\left\vert r\right\vert (1-\left\vert u(t)\right\vert ^{2})^{2}}, \\
c & = & \frac{\tanh ^{2}\left\vert r\right\vert (1-\left\vert u(t)\right\vert ^{2})\left\vert u(t)\right\vert ^{2}}{1-\tanh ^{2}\left\vert r\right\vert
(1-\left\vert u(t)\right\vert ^{2})^{2}}.
\end{eqnarray}


To measure the entanglement in continuous variable system, one generally uses the logarithmic negativity \cite{Werner02}.  The logarithmic negativity of
a bipartite system was introduced originally as
\begin{equation}
E_{N} = \log_2\sum_i\left\vert \lambda^-_i\right\vert,
\end{equation}
where $\lambda^-_i$ is the negative eigenvalue of $\rho^{T_i}$, and $\rho^{T_i}$ is a partial transpose of the bipartite state $\rho$ with respect to
the degrees of freedom of the $i$-th party.  This measure is based on the Peres-Horodecki criterion \cite{Peres96,Horodecki96} that a bipartite quantum
state is separable if and only if its partially transposed state is still positive.

For the continuous-variable (Gaussian-type) bipartite state, its density matrix is characterized by the covariance matrix defined as the second moments
of the quadrature vector $X = (x_1, p_1, x_2, p_2)$,
\begin{equation}
V_{ij} = \frac{\langle\Delta X_i\Delta X_j + \Delta X_j\Delta X_i\rangle}{2},
\end{equation}
where $\Delta X_i = X_i - \langle X_i\rangle$, and $x_i = \frac{a_i + a^{\dag}_i}{\sqrt{2}}$, $p_i = \frac{a_i - a^{\dag}_i}{i\sqrt{2}}$.  The canonical
commutation relations take the form as $[X_i, X_j] = iU_{ij}$, with $U = \left(
\begin{array}{cc}
J & 0 \\
0 & J
\end{array}
\right)$ and $J = \left(
\begin{array}{cc}
0 & 1 \\
-1 & 0%
\end{array}
\right)$ defining the symplectic structure of the system.  The property of the covariance matrix $V$ is fully determined by its symplectic spectrum
$\nu = (\nu_1, \nu_2)$, with $\pm\nu_i$ ($\nu_i > 0$) the eigenvalues of the matrix: $iUV$.  The uncertainty principle exerts a constraint on $\nu_i$
such that $\nu_i \geqslant \frac{1}{2}$ \cite{Adesso05}. Thus the Peres-Horodecki criterion for the continuous-variable state can be rephrased as
the state being separable if and only if the uncertainty principle, $V + \frac{i}{2}U \geqslant 0$, is still obeyed by the covariance matrix under
the partial transposition with respect to the degrees of freedom of a specific subsystem \cite{Simon00}.  In terms of phase space, the action of
partial transposition amounts to a mirror reflection with respect to one of the canonical variables of the related subsystem.  For instance,
$\tilde{V} = \Lambda V\Lambda$, and $\Lambda = diag(1,1,1,-1)$ is the partial transposition with respect to the second subsystem.  If a Gaussian-type
bipartite state is nonseparable, the covariance matrix $\tilde{V}$ will violate the uncertainty principle and its symplectic spectrum
$\tilde{\nu} = (\tilde{\nu}_1, \tilde{\nu}_2)$ will fail to satisfy the constraint $\tilde{\nu}_i \geqslant \frac{1}{2}$.
The logarithmic negativity is then used to quantify this violation as \cite{Werner02}
\begin{equation}\label{measu}
E_N = \max \{0, -\log_2(2\tilde{\nu}_{\min})\},
\end{equation}
where $\tilde{\nu}_{\min}$ is the smaller one of the two symplectic eigenvalues $\tilde{\nu}_i$.  It is evident from Eq. (\ref{measu}) that, if
$\tilde{V}$ obeys the uncertainty principle, i.e., $\tilde{\nu}_i \geqslant \frac{1}{2}$, then $E_N(\rho) = 0$, namely, the state is separable.
Otherwise, it is entangled.  Therefore, the symplectic eigenvalue $\tilde{\nu}_{\min}$ encodes a qualitative feature of the entanglement for an
arbitrary continuous-variable bipartite state.

With this entanglement measure at hand, we study now the entanglement dynamics of the squeezed-state quantum channel in our model.
From the time-dependent state, the covariance matrix for the optical field
can be calculated straightforwardly,
\begin{equation}
V=\left(
\begin{array}{cccc}
\frac{y(1+d)}{2(1-d)^{2}} & 0 & \frac{a\text{Re}[b]}{x} & \frac{a\text{Im}[b]%
}{x} \\
0 & \frac{y(1+d)}{2(1-d)^{2}} & \frac{a\text{Im}[b]}{x} & \frac{-a\text{Re}%
[b]}{x} \\
\frac{a\text{Re}[b]}{x} & \frac{a\text{Im}[b]}{x} &
\frac{y(1+d)}{2(1-d)^{2}}
& 0 \\
\frac{a\text{Im}[b]}{x} & \frac{-a\text{Re}[b]}{x} & 0 & \frac{y(1+d)}{%
2(1-d)^{2}}%
\end{array}%
\right) ,
\end{equation}%
where $x=[(1-c)^{2}-\left\vert b\right\vert ^{2}]^{2}$, $y=\frac{a}{1-c}$,
and $d=c+\frac{\left\vert b\right\vert ^{2}}{1-c}$. And the logarithmic negativity $E_N(t)$ can also be obtained
exactly from Eq. (\ref{measu}).  It is easy to verify that the initial entanglement is $E_N(0) = \frac{2r}{\ln 2}$.

\section{Numerical results and discussions}
\begin{figure}[tbp]
\scalebox{0.37}{\includegraphics{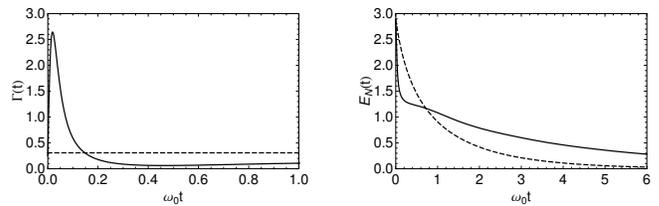}}
\caption{Damping rate $\Gamma(t)$ and logarithmic negativity $E_N(t)$ for the entangled squeezed state as function of dimensionless quantity
$\omega_0 t$ and their corresponding Markovian results (dashed lines).  The parameters $\omega_c/\omega_0 = 50.0$, $\eta = 0.1$, and $r = 1.0$ are used
in the numerical calculation.}\label{f1}
\end{figure}

\begin{figure}[tbp]
\scalebox{0.37}{\includegraphics{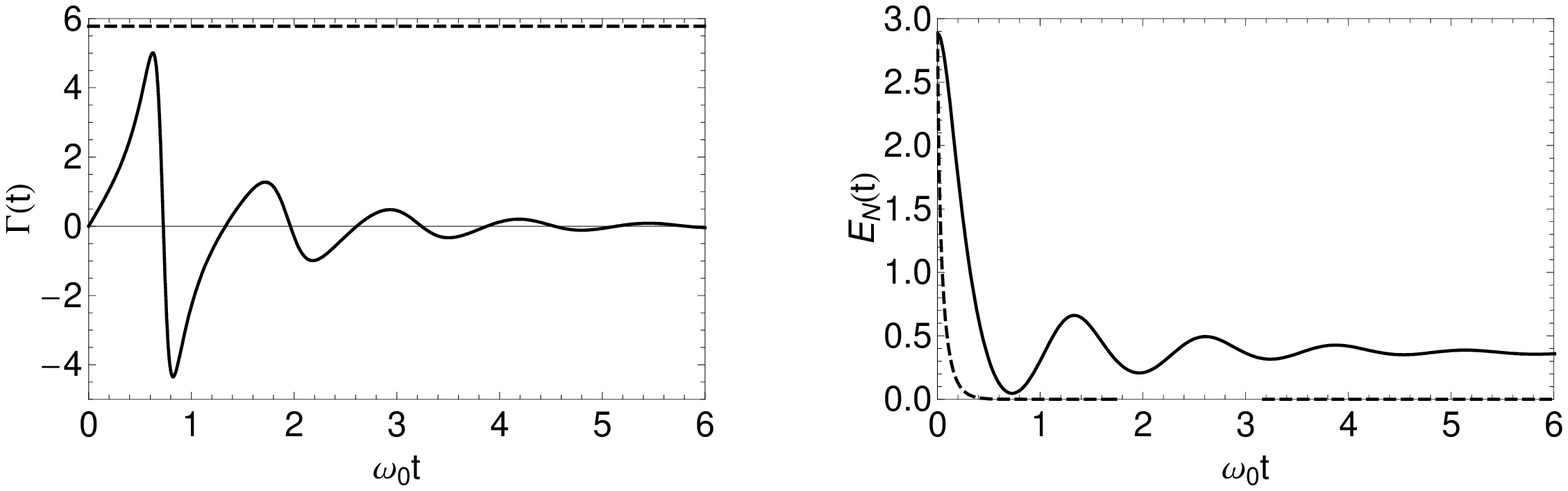}}
\caption{Damping rate $\Gamma(t)$ and logarithmic negativity $E_N(t)$ for the entangled squeezed state as function of dimensionless quantity
$\omega_0 t$ and their corresponding Markovian results (dashed lines).  The parameters $\omega_c/\omega_0 = 1.0$, $\eta = 5.0$, and $r = 1.0$ are used
in the numerical calculation.}\label{f2}
\end{figure}

\begin{figure}[tbp]
\scalebox{0.37}{\includegraphics{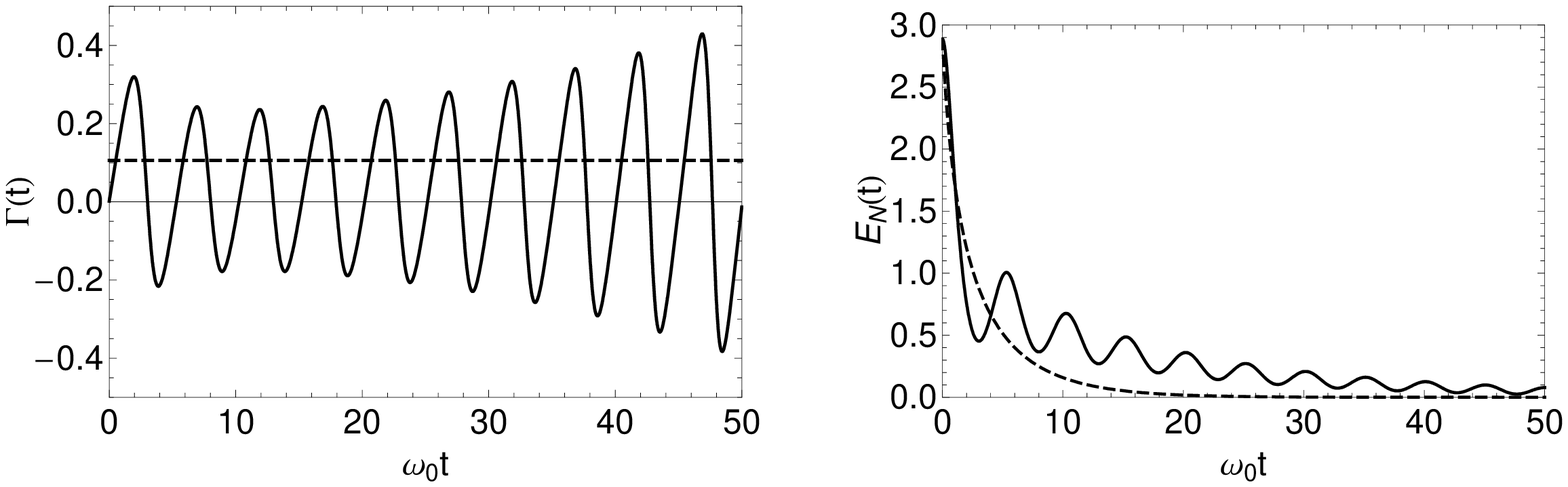}}
\caption{Damping rate $\Gamma(t)$ and logarithmic negativity $E_N(t)$ for the entangled squeezed state as function of dimensionless quantity
$\omega_0 t$ and their corresponding Markovian results (dashed lines).  The parameters $\omega_c/\omega_0 = 0.2$, $\eta = 5.0$, and $r = 1.0$ are used
in the numerical calculation.}\label{f3}
\end{figure}

In the following, we analyze explicitly the exact decoherence dynamics of the entangled squeezed state of $S_1S_2$ under the influence of $E_1E_2$.  For
simplicity, we assume from here on that the two optical fields are identical, i.e., $\omega_1 = \omega_2 \equiv \omega_0$; and they interact with the
same strength, $g_{1l} = g_{2l} \equiv g_l$, with their individual environments.  For definiteness, we consider both $E_1$ and $E_2$ to have Ohmic
spectral density.  The environmental correlation time $\tau_E$ in this case is roughly inversely proportional to the cutoff frequency $\omega_c$ in
Eq. (\ref{spectral}), i.e., $\tau_E \simeq 1/\omega_c$ \cite{Weiss}.  It is emphasized that the cutoff frequency $\omega _{c}$, which is originally
introduced to eliminate infinities in frequency integrations, therefore also determines if the dynamics of open system $S$ is Markovian or non-Markovian.
Our non-perturbatively derived exact results allow us to explore all these possibilities.

In Fig. \ref{f1}, we plot the numerical results of the decay rate $\Gamma(t)$ and logarithmic negativity $E_N(t)$ when $\tau_E \ll \tau_0$.  The
positivity of $\Gamma(t)$ throughout the whole evolution process guarantees the monotonic decrease of $E_N(t)$.  Accordingly, the entangled squeezed
state eventually evolves to a product state, namely the ground state of the system: $\rho_g = |00\rangle\langle00|$.  Clearly, in this case, the
backactions of $E_1E_2$ have a negligible effect on the dynamics of $S_1S_2$, and we say the system dynamics is mainly governed by the dissipative effect
of the environments.  There is thus no qualitative difference between the exact entanglement dynamics and the Markovian results.  Quantitatively,
however, we note that for $t < \tau_E$, the distinctive increase of $\Gamma(t)$ results in $E_N(t)$ decreasing rapidly.  This non-Markovian effect only
shows up in a very short time scale.  In fact, for $t > \tau_E$, $\Gamma(t)$ decreases and approaches gradually to a constant value as $t$ approaches
$\tau_0$; and the rate of decrease of $E_N(t)$ decreases.

Fig. \ref{f2} shows $\Gamma(t)$ and $E_N(t)$ when $\tau_E = \tau_0$.  In this case, the backactions of $E_1E_2$ have a considerable impact on the dynamics
of $S_1S_2$, and the Markovian approximation is not applicable.  Firstly, we note that $\Gamma(t)$ can take negative values.  Physically, this
corresponds to the systems reabsorbing photons from the environments, which leads to an increase in the photon number of the systems \cite{Breuer02}.
These negative decay rates provide evidences for backactions in non-Markovian dynamics \cite{Piilo08}.  Secondly, we observe that $\Gamma(t)$ approaches
zero asymptotically.  Both results clearly differ from the Markovian ones.  Consequently, $E_N(t)$ presnts distinctive behaviors that are absent in the
Markovian results.  Firstly, due to the negative decay rates, $E_N(t)$ shows oscillations.  We must emphasize that these oscillations are fundamentally
different from the transient entanglement oscillations previously obtained when two optical-fields interact with a common environment \cite{An07}.  They
are caused by the backactions of the environments on their respective local optical fields and are characteristic of non-Markovian dynamics.  Similar
oscillations have been obtained in a system of two two-level atoms in two separated damping cavities \cite{Bellomo07}.  Secondly, and more interestingly,
we find that there is some residual entanglement left in the steady state.  From previous studies \cite{Ban06, Goan07, An07}, one would have concluded
that non-Markovian effects only show up in short-time dynamics.  Our results, however, clearly show on the contrary that non-Markovian effects can also
have an influence on the long-time behavior of the system dynamics and the final steady state of the system.  This counterintuitive behavior can be
explained with the fact that the dissipative influence on the entanglement dynamics by the environments is strongly counteracted by the effect due to
their backactions.  Consequently, the decay of the entanglement ceases when the system evolves to some steady state, which is not the ground state
$\rho_g$.

The results when $\tau_E \gg \tau_0$ are shown in Fig. \ref{f3}; a situation considered in Ref. \cite{Paris07}.  Due to extremely long memory of the
environments, the backactions on the systems are so strong that they govern the decoherence dynamics.  As a result, $\Gamma(t)$ and hence $E_N(t)$
oscillate over a very long duration.  These oscillations persist even as the state approaches the ground state.  The `equilibrium' position for the
oscillation of $\Gamma(t)$ is not at zero, but a small positive value.  This positivity means the systems dynamics will experience a weak dissipation,
which is verified by the time evolution of $E_N(t)$ in Fig. \ref{f3}.

In summary, we have studied the exact entanglement dynamics of the two optical-field system for different $\tau_E$'s in comparison with $\tau_0$.
Specifically, we have analyzed when the system dynamics will exhibit novel non-Markovian effects, and provided a detailed description of these.

\section{Conclusions}
We have applied the influence-functional method of Feynman and Vernon to investigate the exact entanglement dynamics of two single-mode optical fields
$S_1S_2$ coupled to two independent and uncorrelated environments $E_1E_2$.  From our analytical and numerical results, it is seen that $E_1E_2$ exert
two competing influences on our system.  One effect, ${\cal D}$, is dissipative and is responsible for the decoherence of $S_1S_2$.  The other,
${\cal B}$, is due to the backactions of $E_1E_2$ on $S_1S_2$.  The degree of manifestations of ${\cal D}$ and ${\cal B}$ in the dynamics of $S_1S_2$
depends on $\tau_E$ in comparison with $\tau_0$.  For $\tau_E \ll \tau_0$, ${\cal D}$ dominates and ${\cal B}$ only gives rise to a transient coherent
oscillation of $S_1S_2$.  The state of $S_1S_2$ evolves to the ground state $\rho_g$, which is coincident with the Markovian result.  If
$\tau_E = \tau_0$, the near resonant interaction between $S_1S_2$ and  $E_1E_2$ results in ${\cal D}$ and ${\cal B}$ being comparable and counteract each
other.  These give rise to transient negative decay rates and asymptotical zero decay rate.  The state of $S_1S_2$ thus evolves asymptotically to some
steady state, which is not the ground state $\rho_g$.  Finally, when $\tau_E \gg \tau_0$, ${\cal B}$ dominates and governs the dynamics of $S_1S_2$.  The
decay rates of the system oscillate about some non-negative equilibrium position over a very long duration.  This slight positivity guarantees an
overall weak dissipative effect on the system dynamics.  Therefore, the state of $S_1S_2$ eventually approaches the ground state with the entanglement
oscillation persisting on for a very long time.

The theory we have established is a non-perturbative description of the exact decoherence dynamics of a system of two single-mode optical fields.  It
can serve as a useful basic theoretical model in analyzing the non-Markovian decoherence dynamics of optical fields employed in practical quantum
information processing schemes.  It should be noted that although only the Ohmic spectral density is considered here, it is straightforward to
generalize our discussion to the non-Ohmic cases.

\section*{Acknowledgement}
The work is supported by NUS Research Grant No. R-144-000-189-305.  J.H.A. also thanks the financial support of the NNSF of China under Grant
No. 10604025, and the Fundamental Research Fund for Physics and Mathematics of Lanzhou University under Grant No. Lzu05-02.


\begin{thebibliography}{99}
\bibitem{Nielsen00} M. A. Nielsen and I. L. Chuang, \textit{Quantum Computation and Quantum Information} (Cambridge University Press, Cambridge, U.K., 2000).
\bibitem{Bennett} C. H. Bennett, G. Brassard, C. Crepeau, R. Jozsa, A. Peres, and W. K. Wootters, Phys. Rev. Lett. {\bf 70}, 1895 (1993).
\bibitem{Bouwmeester97} D. Bouwmeester, J.-W. Pan, K. Matter, M. Eibl, H. Weinfurter, and A. Zeilinger, Nature \textbf{390}, 575 (1997).
\bibitem{Braunstein98} S. L. Braunstein and H. J. Kimble, Phys. Rev. Lett. \textbf{80}, 869 (1998).
\bibitem{Furusawa98} A. Furusawa, J. L. Sorensen, S. L. Braunstein, C. A. Fuchs, H. J. Kimble, and E. S. Polzik, Science \textbf{282}, 706 (1998).

\bibitem{Jakub04} J. S. Prauzner-Bechcicki, J. Phys. A: Math. Gen. \textbf{37}, L173 (2004).
\bibitem{An05} J.-H. An, S.-J. Wang, and H.-G. Luo, J. Phys. A: Math. Gen. \textbf{38}, 3579 (2005).
\bibitem{Rossi06} R. Rossi Jr., A. R. Bosco de Magalh\~{a}es, M. C. Nemes, Physica A \textbf{365}, 402 (2006).

\bibitem{Ban06} M. Ban, J. Phy. A: Math. Gen. \textbf{39}, 1927 (2006).
\bibitem{Goan07} K.-L. Liu and H.-S. Goan, Phys. Rev. A \textbf{76}, 022312 (2007).
\bibitem{Paris07} S. Maniscalco, S. Olivares, and M. G. A. Paris, Phys. Rev. A \textbf{75}, 062119 (2007).
\bibitem{An07} J.-H. An and W. M. Zhang, Phys. Rev. A \textbf{76}, 042127 (2007).

\bibitem{Carmichael93} H. J. Carmichael, \textit{An Open Systems Approach to Quantum Optics}, Lecture Notes in Physics, Vol. m18 (Springer-Verlag, Berlin, 1993).
\bibitem{Breuer02} H.-P. Breuer and F. Petruccione, \textit{The theory of open quantum systems} (Oxford University Press, Oxford, 2002).

\bibitem{Dubin07} F. Dubin, D. Rotter, M. Mukherjee, C. Russo, J. Eschner, and R. Blatt, Phys. Rev. Lett. \textbf{98}, 183003 (2007).
\bibitem{Koppens07} F. H. L. Koppens, D. Klauser, W. A. Coish, K. C. Nowack, L. P. Kouwenhoven, D. Loss, and L. M. K. Vandersypen, Phys. Rev. Lett. \textbf{99}, 106803 (2007).
\bibitem{Mogilevtsev08} D. Mogilevtsev, A. P. Nisovtsev, S. Kilin, S. B. Cavalcanti, H. S. Brandi, and L. E. Oliveira, Phys. Rev. Lett. \textbf{100}, 017401 (2008).

\bibitem{Weiss} U. Weiss, \textit{Quantum Dissipative Systems}, 2nd ed. (World Scientific, Singapore, 1999).

\bibitem{Feynman63} R. P. Feynman and F. L. Vernon, Ann. Phys. (N. Y.) \textbf{24}, 118 (1963).
\bibitem{Leggett87} A. J. Leggett, S. Chakravarty, A. T. Dorsey, M. P. A. Fisher, A. Garg, and W. Zwerger, Rev. Mod. Phys. \textbf{59}, 1 (1987).

\bibitem{An072} J.-H. An, M. Feng, W. M. Zhang, arXiv:0705.2472v2 [quant-ph].
\bibitem{An08} J.-H. An, Y. Yeo, and C. H. Oh, arXiv:0808.3178v1 [quant-ph].

\bibitem{Zhang90} W. M. Zhang, D. H. Feng, and R. Gilmore, Rev. Mod. Phys. 62, 867 (1990).

\bibitem{Hu92} B. L. Hu, J. P. Paz, and Y. Zhang, Phys. Rev. D \textbf{45}, 2843 (1992).

\bibitem{EPR35} A. Einstein, B. Podolsky, and N. Rosen, Phys. Rev. \textbf{47}, 777 (1935).

\bibitem{Ou92} Z. Y. Ou, S. F. Pereira, H. J. Kimble, and K. C. Peng, Phys. Rev. Lett. \textbf{68}, 3663 (1992).

\bibitem{Pielawa07} S. Pielawa, G. Morigi, D. Vitali, and L. Davidovich, Phys. Rev. Lett. \textbf{98}, 240401 (2007).

\bibitem{Werner02} G. Vidal and R. F. Werner, Phys. Rev. A \textbf{65}, 032314 (2002).

\bibitem{Peres96} A. Peres, Phys. Rev. Lett. \textbf{77}, 1413 (1996).
\bibitem{Horodecki96} M. Horodecki, P. Horodecki, and R. Horodecki, Phys. Lett. A \textbf{223}, 1 (1996).

\bibitem{Adesso05} G. Adesso and F. Illuminati, Phys. Rev. A \textbf{72}, 032334 (2005).

\bibitem{Simon00} R. Simon, Phys. Rev. Lett. \textbf{84}, 2726 (2000).

\bibitem{Bellomo07} B. Bellomo, R. LoFranco, and G. Compagno, Phys. Rev. Lett. \textbf{99}, 160502 (2007).

\bibitem{Piilo08} J. Piilo, S. Maniscalco, K. H\"{a}rk\"{o}nen, and K-A. Suominen, Phys. Rev. Lett. \textbf{100}, 180402 (2008).

\end{thebibliography}
\end{document}